# Specific features of the dielectric continuum solvation model with a position-dependent permittivity function.

*M.V. Basilevsky [1*], F. V. Grigoriev[2,3], O. Yu. Kupervasser[2,3]*

[1] Photochemistry center, Russian Academy of Sciences , Russia, 119421 Moscow, ul. Novatorov, d. 7a;

[2] Research Computing Center M.V. Lomonosov Moscow State University, Russia , 119992 Moscow, Vorobjovi Gory, MGU NIVC;

[3] Dimonta, Ltd., 117186, 15 Nagornaya Str., bldg 8, Moscow, Russia.

AUTHOR EMAIL ADDRESS: M.V. Basilevsky, E-mail: basil@photonics.ru; F. V. Grigoriev, E-mail: fedor.grigoriev@gmail.com; O. Yu. Kupervasser, olegkup@yahoo.com

CORRESPONDING AUTHOR FOOTNOTE: F. V. Grigoriev, Research Computing Center M.V. Lomonosov Moscow State University, Russia , 119992 Moscow, Vorobjovi Gory, MGU NIVC, E-mail: fedor.grigoriev@gmail.com;

**Abstract**

We consider a modified formulation for the recently developed new approach in the continuum solvation theory (Basilevsky, M. V., Grigoriev, F. V., Nikitina, E. A., Leszczynski, J. *J. Phys. Chem. B* **2010,** *114,* 2457), which is based on the exact solution of the electrostatic Poisson equation with the space–dependent dielectric permittivity. Its present modification ensures the property curl $E = 0$ for the electric strength field $E$ inherent to this solution, which is the obligatory condition imposed by Maxwell equations. The illustrative computation is made for the model system of the point dipole immersed in a spherical cavity of excluded volume.

1. **Introduction and notation.**

The new efficient algorithm providing the exact solution for the Poisson electrostatic equation with a space–dependent dielectric permittivity function $\varepsilon(r)$ (where $r$ is a space point vector) has been formulated recently [1]. It proved to be a useful tool for applications in the solvation theory. The objective of the present note is to refine the derivation of its underlying equations without changing the ultimate results and the computational scheme.

The necessary preliminary notations are introduced below. So, $E = -\nabla \psi$ represents the electric field strength $E$ and the electrostatic potential $\psi$; $E_0 = -\nabla \psi_0$ corresponds to the vacuum electric field strength $E_0$ (where $\varepsilon = 1$) and the pertaining vacuum potential $\psi_0$; the vacuum field obeys the Poisson equation $\nabla^2 \psi_0 = -4\pi\rho$, where $\rho$ is the charge distribution of a solute. The displacement field is, at the primary step, defined conventionally as

$$D = \varepsilon E \qquad (1)$$

The background static Maxwell equations read:

$$\nabla D = 4\pi\rho \qquad (a)$$

$$\nabla \times E = 0 \qquad (b) \qquad (2)$$

It was shown [1] that the exact solution of Eq. (2a) can be expressed in terms of the $D - E_0$ theorem [2-4]:

$$D = E_0 \qquad (3)$$

In [1] the solute immersed in the non-uniform continuum medium with a position–dependent dielectric permittivity function $\varepsilon(r)$ has been considered. As usually, the solute was contained in a cavity of excluded volume where $\varepsilon = 1$. Eq 3 is valid for the case of a cavity with complicated shape, provided the function $\varepsilon(r)$ is continuous everywhere in space. Based on this idea the explicit algorithm performing a computation of solvation effects was elaborated [1]. The computational scheme, called SBCM (smooth boundary continuum model), proved to work successfully for various applications. Two versions of this methodology (SBCM-1 and SBCM-2) were based on the following expressions for the total electrostatic free energy $G_{el}$ and the solvation free energy $G_{solv}$:

$$G_{el} = \frac{1}{8\pi}\int ED d^3r = \frac{1}{8\pi}\int \frac{1}{\varepsilon}(\nabla \psi_0)^2 d^3r$$

$$G_{solv} = -\frac{1}{8\pi}\int \left(1 - \frac{1}{\varepsilon}\right)(\nabla \psi_0)^2 d^3r \qquad (4)$$

(SBCM-1)

$$G_{solv} = \frac{1}{2}\int \rho \Phi d^3 r$$

$$\Phi(r) = \int \frac{g(r')}{|r-r'|} d^3 r'; \quad g(r) = \frac{1}{4\pi\varepsilon^2}(\nabla\varepsilon\nabla\psi_0) \qquad (5)$$

(SBCM-2)

In eq 5 $\Phi(r)$ is the response field defined as $\psi = \psi_0 + \Phi$ and $g(r)$ is the induced polarization charge. This equation was derived from the generalized (including the medium) Poisson equation $\nabla(\varepsilon\nabla\psi) = -4\pi\rho$ which is equivalent to eq 2a when eq 1 is valid.

According to eqs 1 and 3, $E$ was considered [1] as

$$E = -\frac{1}{\varepsilon}\nabla\psi_0 \qquad (6)$$

The inconsistency accompanying this step is revealed after $\nabla\times E$ is calculated:

$$\nabla\times(-\frac{1}{\varepsilon}\nabla\psi_0) = \frac{1}{\varepsilon^2}(\nabla\varepsilon\times\nabla\psi_0) \qquad (7)$$

Provided $\nabla\varepsilon \neq 0$, which is the main objective addressed here, the right hand part of eq 7 vanishes only in few exceptional cases [1-3]. As a result, eq 2b is violated.

This inconsistency is explained and resolved in the present work, thus clearly establishing the status of the SBCM approach. Sections 2 and 3 are devoted to the general consideration of the question. As a particular example, we formulate in section 4 the problem of solvation for a point dipole in a spherical cavity with the continuous position-dependent dielectric function $\varepsilon(r)$ in the external region. This is an extension of the well – known classic Onzager dipole model, in which $\varepsilon = \varepsilon_0$ is treated as a constant everywhere outside the cavity. The full solution for the case of variable $\varepsilon(r)$ requires sophisticated computations even for this extremely simple system. Their results are reported and discussed in the main text (sections 5 and 6), whereas the technical computational details are transferred to the four Appendices A-D.

2. **The solid background for the SBCM**

In order to satisfy eq 2b, the definition of the vector field $E$ suggested by eq 6 must be modified. Using the Helmholtz representation [5] for this expression in terms of its gradient ($E_1$) and curl ($E_2$) components we find the partitioning scheme:

$$-\frac{1}{\varepsilon}\nabla\psi_0 = E_1 + E_2$$
$$E_1 = -\nabla\psi; \ E_2 = \nabla \times A$$
(8)

Here the scalar (electrostatic) potential $\psi$ and the vector potential $A$ are introduced. They are defined as [5]:

$$\psi(r) = \frac{1}{4\pi}\int \frac{1}{|r-r'|}\nabla\left(-\frac{1}{\varepsilon}\nabla\psi_0\right)d^3r' = \int \frac{Q(r')}{|r-r'|}d^3r'$$

$$A(r) = \frac{1}{4\pi}\int \frac{1}{|r-r'|}\nabla\times\left(-\frac{1}{\varepsilon}\nabla\psi_0\right)d^3r' = \int \frac{I(r')}{|r-r'|}d^3r'$$
(9)

For clarity, the derivation of eq 9 is extended in Appendix A.

In this way the scalar charge $Q(r')$ and the vector triplet of sources $I(r')$ are determined. Their direct evaluation yields

$$Q(r') = \frac{1}{4\pi}\left(\frac{1}{\varepsilon^2}\nabla\varepsilon\nabla\psi_0 - \frac{1}{\varepsilon}\nabla^2\psi_0\right) = g(r') + \frac{1}{\varepsilon}\rho(r')$$
(10)

$$I(r') = \frac{1}{4\pi\varepsilon^2} \nabla\varepsilon \times \nabla\psi_0 \qquad (11)$$

In eq 10 $g(r')$ is the polarization charge defined in eq 5 and $\rho(r')$ is the solute charge density. The vector source $I(r')$ in eq 11 is identical with the curl 7. It is also seen that the first line of eq 9 together with eq 10 exactly reproduce the SBCM-2 computational scheme for the solvation energy (i. e. the eq 5) as devised earlier [1]. So eq 5 for the response field $\Phi(r)$ appears under the condition that all solute charges are contained inside the cavity where $\varepsilon = 1$ (i. e. $(\varepsilon -1)\rho = 0$ [1,6]).

Based on these results, we can now modify the definition of the electric field strength $E$ as

$$E = E_1 = -\frac{1}{\varepsilon}\nabla\psi_0 - E_2 \qquad (12)$$

It obviously satisfies eq 2b and suggests the computational scheme for the solvation energy, which was already established as SBCM-2.

The solenoidal term $E_2$ arises as a spurious component of $E$ when this field is inconsistently represented in the form of eq 6. Its appearance seemingly distinguishes the SBCM-1 and SBCM-2 algorithms. For the electrostatic energy the following misfit quantity is obtained:

$$\delta G = G_{el}(SBCM-1) - G_{el}(SBCM-2) = \frac{1}{8\pi}\int E_2 D d^3r =$$
$$= -\frac{1}{8\pi}\int \nabla\psi_0(r)[\nabla \times A(r)]d^3r$$

The use of the identity $\nabla\psi_0[\nabla \times A(r)] = \nabla[A(r) \times \nabla\psi_0]$ transforms this equation into

$$\delta G = -\frac{1}{8\pi}\int_V \nabla \bullet [A(r) \times \nabla \psi_0] d^3 r = -\frac{1}{8\pi}\int_S n \bullet [A(r) \times \nabla \psi] dS$$

The volume integral is transformed into the surface one where the closed surface $S$ surrounds the volume $V$ and $n$ is the unit normal direction at $dS$. Similar to the derivation in Appendix A, the surface can be shifted far away from the solute region where $\nabla \varepsilon$ vanishes and $A(r) = 0$. In this way we obtain $\delta G = 0$ and $G_{el}(SBCM-1) = G_{el}(SBCM-2)$. The similar equality is valid for the solvation free energies.

### 3. Two alternative interpretations of Eq 12.

The first interpretation is most straightforward. Under the condition $D = \varepsilon E$ (eq 1) eq 12 is considered as an approximation to the exact solution of eq 2a. The essence of this approximation is revealed as follows. By applying eq 1 we convert eq 12 firstly into $D = -\nabla \psi_0 - \varepsilon E_2$ and next into $\nabla D = 4\pi(\rho - \delta\rho)$ where

$$\delta\rho = \frac{1}{4\pi}\nabla(\varepsilon E_2) \qquad (13)$$

The comparison with Eq 2a signals that $\delta\rho$ is the spurious charge, a measure of the pertaining inaccuracy. It is explicitly evaluated in Appendix B. The way to the accurate treatment is traced by introducing the exact potential,

$$\Psi = \psi + \varphi \qquad (14)$$

and the corresponding exact electric field strength

$$E = -\frac{1}{\varepsilon}\nabla\psi_0 - E_2 - \nabla\varphi = E_1 - \nabla\varphi \qquad (15)$$

Thereby, $\nabla\varphi$ serves as the desired correction to eq 12. Based on eq 15 the reevaluation of $D$ and $\nabla D$ regenerates eq 2a under the condition

$$\nabla(\varepsilon\nabla\varphi) = -4\pi\delta\rho \qquad (16)$$

This Poisson-like equation defines the correction potential $\varphi$. The discussion of its solution is transferred to Appendix B.

The alternative second interpretation looks more tricky. Let us assume that the new pair of fields $D$ (eq 3) and $E$ (eq 12) obey Maxwell equations 2 but the conventional linear connection rule formulated as eq 1 is invalid. As a concluding step, it must be consistently changed in order to become compatible with the new eq 12. The desired connection reads:

$$E = \frac{1}{\varepsilon}D - \frac{1}{4\pi}\nabla \times \int \frac{1}{\varepsilon^2(r')}(\nabla' \varepsilon(r') \times D(r'))\frac{1}{|r-r'|}d^3r' \qquad (17)$$

The conventionally contracted form of this expression uses the integral kernel $\frac{1}{\varepsilon(r,r')}$ as defined below:

$$E = \int \frac{1}{\varepsilon(r,r')}D(r')d^3r'$$
$$\frac{1}{\varepsilon(r,r')} = \frac{1}{\varepsilon(r')}\delta(r-r') + \frac{1}{4\pi}(\nabla \frac{1}{|r-r'|}) \times [\nabla' \frac{1}{\varepsilon(r')} \times ...] \qquad (18)$$

Here and henceforth we distinguish the gradient operators $\nabla$ and $\nabla'$ as those acting on the variables $r$ and $r'$ respectively.

Eq 18 represents a typical non-local expression [3,7,8] with the kernel $\varepsilon^{-1}(r,r')$. Its non-local tensorial second part serves as a correction annihilating the spurious solenoidal fraction of the operator $\frac{1}{\varepsilon}\nabla$. In this way, the combination of eqs 3,12,17,18 provides the exact solution for problem 2 with a non-local relation 18 interconnecting the $E$ and $D$ fields. The SBCM-2 prescription 5 remains with no changes as the working algorithm which underlies the practical implementation of this solution. As proved in section 2, the free energy eqs 4 and 5 remain exact and equivalent.

The non-local conditions 17,18 can be reformulated in a different but entirely equivalent form, which follows from eqs 8,9 for $E_1$ (which defines $E$ according to eq 12):

$$E(r) = \nabla \left\{ \frac{1}{4\pi} \int \frac{d^3 r'}{|r-r'|} \left[ \frac{1}{\varepsilon(r')} \nabla' + \nabla' \frac{1}{\varepsilon(r')} \right] D(r') \right\} \quad (19)$$

The expression in curly brackets represents the potential $\psi$ which generates the exact field $E$. After substituting $D = -\nabla \psi_0$ ( i. e. the $D-E_0$ theorem), it turns into the SBCM-2 potential, as given by eqs 9,10. Being equivalent to eq 17, the eq 19 suggests a much simpler algorithm, because the tensorial nature of the algorithm 18, introduced via the vector product "×", is eliminated in terms of the new transcription. The $E-D$ relation 19 together with the pair of Maxwell equations 2 provides the most compact formulation of the SBCM as the rigorous non-local procedure.

**4. The sample case: the Onsager point dipole with smooth boundary.**

In the present section we consider an approximate evaluation of the difference between the exact free energy $G_{el}$, extracted from the solution of Eqs (1,2) (without the approximation 3), and the SBCM energies (4) and (5) (they are rigorously equivalent, as shown at the end of section 2). The computation is performed for the point dipole with moment $m$ positioned at the centre of the spherical cavity of radius $a$ with $\varepsilon = 1$ inside the cavity. The detailed vectorial notation will be used henceforth for the space variables, namely $\vec{r}(r, \vartheta, \varphi)$ and $\vec{r}'(r', \vartheta', \varphi')$, with $r, \vartheta, \varphi$ and the primed counterparts being spherical coordinates. Notations $d^3 r$ and $d^3 r'$ are retained for volume differentials. The centre of the cavity is placed at the origin of the coordinate frame. The spherically symmetric dielectric function depends on the radius $r$ [1] as

$$\varepsilon(r) = 1 + 4\pi \chi z(r)$$
$$z(r) = 1 + \exp[-2\alpha(r-a)] - 2\exp[-\alpha(r-a)] \quad (r > a) \quad (20)$$
$$z(r) = 0 \quad (r < a)$$

Here $\chi$ is the standard dielectric susceptibility and $z(r)$ represents the dimensionless solvent density. Its growth begins at the cavity boundary and the asymptotic value $z = 1$ is reached when $\alpha(r-a) \gg 1$, converging to the static permittivity value $\varepsilon(r) = \varepsilon_0$. Parameter $\alpha$ measures the steepness of the density evolution.

The free energy correction to the SBCM result (4) or (5) will be denoted as $\delta G_{el}$. It originates from the interaction between the Onsager dipole and the polarization charge density $\delta\rho$, which generates the response potential $\varphi$ calculated below as

$$\varphi(\vec{r}) = \varphi_0(\vec{r}) + \varphi_1(\vec{r})$$
$$\varphi_0(\vec{r}) = \int \frac{d^3 r'}{|\vec{r}-\vec{r}'|} \frac{\delta\rho(\vec{r}')}{\varepsilon(\vec{r}')} \qquad (21)$$
$$\varphi_1(\vec{r}) = \int \frac{d^3 r'}{|\vec{r}-\vec{r}'|} \frac{\nabla' \varepsilon(\vec{r}') \nabla' \varphi_0(\vec{r}')}{\varepsilon(\vec{r}')}$$

The third line represents the perturbational version of Eq B6 ($\varphi_0$ in (21) stands for $\varphi$ in (B6)).

The explicit expressions for $\delta\rho$ (eq 13) and $\varphi_0$, $\varphi_1$ (eq 21) are derived in Appendices B and C. Provided the Onsager dipole is arranged along $z$-axis, the desired free energy excess is

$$\delta G_{el} = \frac{1}{2} m^2 \frac{d\varphi}{dz}\bigg|_{r=0} \qquad (22)$$

This computation is not a simple task. The accurate evaluation of the response potential $\varphi$ requires a preliminary exact solution of eq 16. This prescription is simplified in eq 21, where the first step of the iteration procedure described in Appendix B is applied. The details of such computation are discussed in Appendices C and D. Here we explain only the most approximate version of the final result, in which the angular dependence of the polarization density $\delta\rho(\vec{r})$ is simplified significantly (see Eq C3). It reads

$$\delta G_{el} = -\frac{2}{9} m^2 (A + \frac{1}{3} B)$$
$$A = \int_a^\infty dr \int_a^\infty dr' K_0(r, r') \frac{1}{\varepsilon(r')} \frac{d\varepsilon(r)}{dr} \frac{d}{dr'}(-\frac{1}{\varepsilon(r')}) \qquad (23)$$

Here $A$ and $B$ represent the contributions coming from the potentials $\varphi_0$ and $\varphi_1$ in Eq 21. Two quadratures over two radial variables $r$ and $r'$ are involved in $A$. The explicit expression for $B$, containing three such quadratures, can also be found in Appendix C (see eq 7). The kernel $K_0(r,r')$ in eq 23 represents the result of approximate integration of the true kernel $1/|\vec{r}-\vec{r}'|^3$ over spherical angles. Its explicit expression is given as eq C3 (see Appendix C) and motivated in Appendix D.

Applying this simplification in multiple integration steps, which are involved implicitly in (21), leads to the ultimate eq. 23 with no further approximations.

## 5. Computations

The illustrative computations were performed within the following strategy. At the first step the zero order potential $\varphi_0(\vec{r})$ in eq 21 has been calculated using the rigorous algorithm (eqs C1 and C2). It is responsible for the pertaining zero-order fraction of the free energy contribution appearing in terms of eq 22 and denoted as $\delta G_{el}^{(0)}$. Five quadratures (two radial ones and three angular ones) are required for such computation. The similar calculation was repeated based on the approximation 23; the result being

$$\delta G_{el}^{(0)} \approx -\frac{2}{9} m^2 A \qquad (24)$$

Two radial quadratures are involved in this case. Typical results of the exact and approximate calculations are compared in Fig 1 and discussed below.

At the second step we calculated the first order correction potential $\varphi_1(\vec{r})$ and the corresponding part of the free energy misfit, denoted as $\delta G_{el}^{(1)}$. The approximation 23 yields the result:

$$\delta G_{el}^{(1)} \approx -\frac{2}{27}m^2 B \qquad (25)$$

Three radial integrations are actually involved in eq 25. The relative importance of this correction can be estimated from Fig 2.

As discussed in Appendix D, the approximate computations according to eqs 23-25 include the fitting parameter $\delta$, which is inserted in the integral kernel $K_0(r,r')$, i.e. the factor entering the integrand in eq 23 which defines $A$. This parameter depends on the asymptotic bulk value $\varepsilon_0 = 1 + 4\pi\chi$ of the permittivity $\varepsilon(r)$ (see (20)), and it is determined in such a way that the approximate perturbational expression 23 reproduced exactly the non – perturbational rigorous stepwise limit of the total misfit $\delta G_{el}$; this limit is found analytically [1] (see eq D7). Only the zero-order term $\delta G_{el}^{(0)}$ is displayed in Fig. 1. Its exact computation (profile I, eqs C1 and C2) involves no approximations and adjustable parameters. The alternative approximate computation of this term (profile II, eq 24) is based on the above-mentioned empirical parameterization of the kernel $K_0(r,r')$. As found by means of eq D9, $\delta = 0.43$ for $\varepsilon_0 = 20$. The two computations are entirely independent. The fair agreement between these two zero-order profiles verifies the validity of the approximation C3 and its calibration as described in Appendix D. This approximate procedure provides the exact asymptotic ($\alpha \to \infty$) value of the total misfit. As is seen from Fig. 1 for the zero-order term, its extrapolation to low $\alpha$ - values reproduces well the rigorous computation with no special fitting parameters.

It seems that the flat minimum of the exact profile I in Fig. 1 results from the computational artifact. It arises owing to the accumulation of numerical errors during the rigorous computation (it involves five quadratures) when $\alpha$ becomes large. Thereby, the practical advantage of the approximate model (two quadratures for $\delta G_{el}^{(0)}$ and three ones for $\delta G_{el}^{(1)}$) is revealed, as this approach tends smoothly to the asymptotic (stepwise) limit of $\delta G_{el}$ and reproduces it accurately owing to the proper choice of $\delta(\varepsilon_0)$.

We can finally define the dimensionless parameter

$$\gamma = \left| \frac{\delta G_{el}}{G_{solv}(SBCM)} \right| \qquad (26)$$

where the misfit value is evaluated according to the approximate perturbation approach (eqs. 23-25). This is the appropriate measure of the discrepancy between the SBCM and the alternative complete treatment of Maxwell equations 2 with the conventional non-local material relation (eq 1). For the present dipole system, $\gamma$ does not depend on the value of the dipole moment $m$ because both the numerator and denominator in (26) are proportional to $m^2$.

Plots of $\gamma(\alpha)$ are shown in Figs 3 and 4 for several values of $\varepsilon_0$. They all converge to the asymptotic value $\gamma \cong 1/2$ in the step limit $\alpha \to \infty$. The most remarkable observation is that $\gamma(\alpha)$ does not vanish when $\alpha \to 0$. This result follows from the structure of the underlying equations (the both ingredients in (26) tend to zero in this limit) and it leads to the conjecture which is important for applications. The distinction between the two solvation models, as suggested and discussed in section 3 (i.e. those based on the local and the non-local $E$-$D$ relations), is measured by the ratio

$$\frac{G_{solv}(SBCM)}{G_{solv}(SBCM) + \delta G_{el}} = \frac{1}{1+\gamma} \qquad (27)$$

This estimate of the discrepancy is valid within the total $\alpha$ range. The pertaining exact expression for $\gamma$ is are readily available for the step limit $\alpha \to \infty$ (see Appendix D). For large values of the bulk permittivity $\varepsilon_0$ it provides a transparent illustration of the matter under present discussion. When $\varepsilon_0 > 10$, eq D7 can safely be simplified as

$$\delta G_{el} \approx -\frac{m^2}{6a^3}\left(1-\frac{1}{\varepsilon_0}\right) = \frac{1}{2}G_{solv}(SBCM)$$ (the exact stepwise estimate of the SBCM free energy

as the right hand part of this equation which was obtained earlier [1]). Thereby $\gamma = 1/2$ and

$$1 > \frac{1}{1+\gamma} > \frac{2}{3} \qquad (28)$$

Because the exact asymptotic result D7 for $\delta G_{el}$ goes down for lower values of $\varepsilon_0$, the range (28) is conserved in the step limit throughout the total permittivity range, including its small values.

The inspection of Figs 3 and 4 tells us that the estimates specific for the step limit don't change significantly in the wide $\alpha$-range. We infer therefore that the values within the range $\frac{1}{1+\gamma} = 0.6 \div 0.8$ are reliable when $0.5 < \alpha < 3$ Å$^{-1}$, i.e. in the interval of $\alpha$ which is typical for solvents of the real interest [1]. This conclusion is insensitive to the solute size (i.e. to the change of the cavity radius $a$), which obeys rigorously in the step limit (28), whereas a moderate effect of the size variation observed for finite values of $\alpha$ (Fig. 4) does not modify it markedly. More definite statements would be premature at present, remembering that the data for finite $\alpha$ values were computed using the first iteration step of the perturbation approach based on the extra approximation 23 and also keeping in mind the numerical problems which arise in the integrations involved in the present computations when $\alpha$ values become large.

6. **Conclusion.**

The essence of the present study is the SBCM algorithm [1] reformulated in terms of eq 12. Two points of viewing its status are equally well legitimate. According to the first one, eq 12 suggests an approximate computational scheme. This conjecture is based on the standard local relation (1) between $D$ and $E$ fields. The second alternative point of view substitutes this local connection rule by its properly modified non-local counterpart (19) which allows for considering eq 12 as the background for the exact computational scheme.

The first formulation also suggests the route for reaching its exact solution by solving eq 16. The correction $\delta G_{el}$ to the SBCM solvation energy $G_{solv}$(SBCM) then appears, providing the dimensionless smallness parameter $\gamma$, see eq 26. Its exact evaluation constitutes a considerable computational task which is much more difficult than the SBCM procedure itself. Provided this task is performed, the two different algorithms for solving the position-dependent non-uniform $\varepsilon$ problem are totally designed. The magnitude of the parameter $\gamma$ is the measure of the discrepancy between their solutions, which represent two different solvation models. The first local model represents the exact solution for the combination of three equations 1-2, whereas the second one is the SBCM with its interpretation given in terms of the non-local material relation (19). The both algorithms are governed by the same dielectric function $\varepsilon(r)$ (see eq 20) with basic parameters being $\varepsilon_0$ (the static bulk permittivity) and $\alpha$ (the steepness of the variation of $\varepsilon(r)$ on the boundary of the solute cavity). The distinction between them becomes clearly visible in the stepwise limit of the $\varepsilon$-function. Then the first local algorithm in its full version (including the correction according to the eq 16) converges to the classical continuum theory (Born, Kirkwood, Onsager and the recent implementation in terms of the polarizable continuum model (PCM [10])), in which the stepwise $\varepsilon$-approach is introduced as a primary background. The second (non-local) SBCM algorithm reveals a different behavior in this extreme [1]. The so arising discrepancy extends as well for the case of smooth boundary with finite $\alpha$ values, as demonstrated in sections 4 and 5. It is seen now how the two distinguishing solvation algorithms appear when the continuous position–dependent $\varepsilon$ - model is invoked. Such divergence caused earlier a seeming paradox which was not fully resolved in [1].

The numerical comparison of the two models requires a reasonably simple evaluation of eq 26, which is only available for a small highly symmetric system like as Onsager point dipole in a spherical cavity with the abrupt boundary ($\alpha \to \infty$). For this idealized case the estimation $\gamma \approx 1/2$ was obtained [1]. Its extrapolation to finite values of $\alpha$ is made in the present sections 4 and 5 within an approximate perturbational treatment. The basic outcome, represented by eqs 27 and 28, seems to be reliable for real solvents. The computation is system-dependent; moreover, the violation of the $D - E_0$ theorem (eq 3), which is the origin of the discrepancy (27) [2], becomes the most apparent for the present point dipole system, as follows from the examination of its step limit in [1]. For real solutes, in which their charge distributions are not so singular, the lower limit appearing in the right hand part of the inequality 28 is expected to increase, thus decreasing the discrepancy of the two models.

The final comment is addressed to the practical utility of the present methodologies. The distinction between the two models does not affect the quality of any of them in applications. This methodological impact is neutralized by an appropriate parameterization which is different for the two models. The computations of ionic hydration free energies can serve as an illustrative example. Their treatment proved to be quite satisfactory in terms of both PCM [10] and SBCM [1]. In the present context, these two techniques represent the two models under discussion, distinguishing by the choice of their steepness parameter defined as $\alpha \to \infty$ or $\alpha = 3$ Å$^{-1}$ for the PCM (the first model) and for the SBCM (the second model), respectively.

The attempt of treating the non-uniform dielectric medium at a continuum level seems natural. Most transparent is its original formulation in terms of the continuous $\varepsilon(r)$ combined with the local linear relation (1) between $D$ and $E$ fields. Although being physically relevant, this elementary scheme proves to be mathematically controversial without the correction, which implies the task of finding the solution to the eq 16. Its formal purification, given in the present work in terms of the non-local reformulation, may distort the physical content of this model in several cases. When the correction introduced by eq 16 in the local model becomes significant, one may doubt whether the alternative non-local solvation approach, although being mathematically consistent, remains physically adequate. Such a case (with an extremely abrupt boundary of the solute cavity) should be better treated in terms of the alternative and conventional PCM-like methodology [10] with the stepwise $\varepsilon$-function. In particular, the extremely steep change of $\varepsilon$ creates numerical problems at the stage of the gradient calculation in the SBCM-2 as well as in our present computations. They are absent in the stepwise PCM procedure. Therefore, the two different models can serve for describing different sorts of systems. This issue has already been addressed earlier [1] with the reservation that the situation, which is fortunate for the smooth continuum treatment, seems to be typical for the majority of real solution systems, in which both solute and solvent particles are not supposed to be small. Such ultimate conjecture justifies the practically available route for incorporating a position-dependent $\varepsilon$-function in applications of the solvation theory.

## Acknowledgement

The authors are grateful to the Ministry of education and science of Russian Federation for the financial support (the state contract 02.523.11.3014).

**Appendix A. Basic relations underlying the Helmholtz partitioning of vector fields.**

The Helmholtz algorithm formulated in eqs 8,9 is based on the following obvious identities:

$$\nabla E_1 = -\nabla^2 \psi = -\nabla(\frac{1}{\varepsilon}\nabla \psi_0) = 4\pi Q \qquad \text{(A1)}$$

$$\nabla \times E_2 = \nabla \times \nabla \times A = -\nabla \times (\frac{1}{\varepsilon}\nabla \psi_0) = 4\pi I \qquad \text{(A2)}$$

Here the scalar charge density $Q$ and the triplet of vector sources $I$ are completely determined by the direct evaluation performed in eqs 10,11.

The eq 9 for $\psi(r)$ follows immediately from (A1) as a solution to the Poisson equation. The second line in eq 9, addressed for obtaining $A(r)$, can be derived from (A2) by applying the identity $\nabla \times \nabla \times A = \nabla(\nabla A) - \nabla^2 A$. It follows then that the result

$$A(r) = \int \frac{I(r')}{|r-r'|} d^3 r' \qquad (A3)$$

appears under the condition

$$\nabla A = 0 \qquad (A4)$$

Eqs A3,A4 are, indeed, compatible. This is verified by applying the operator div to (A3):

$$\nabla A = \int \nabla \left( \frac{I(r')}{|r-r'|} \right) d^3 r' = -\int I(r') \nabla' \left( \frac{1}{|r-r'|} \right) d^3 r' = -\int \nabla' \left( \frac{I(r')}{|r-r'|} \right) d^3 r' \qquad (A5)$$

As the final step we have applied the fact that $I(r')$ has the structure of a curl, i.e. $\nabla' I(r')=0$. The concluding result appearing in eq A5 is transformed into a surface integral by means of the divergence integral theorem:

$$\nabla A = \int_S \frac{I(r')n(r')}{|r-r'|} dS' = 0 \qquad (A6)$$

where $dS'$ is the element of a closed surface $S'$ and $n(r')$ means the unit normal direction at $dS'$. The surface can be shifted far apart from the solute region, where $\nabla \varepsilon = 0$ and $I(r')$ vanishes.

**Appendix B. Beyond the SBCM solution.**

The conventional local $D-E$ interrelation is applied below as a background. Viewed from this point, the SBCM should be considered as an approximate procedure. It can be refined by solving eq 16 for the correction potential $\varphi$. The essential steps of this treatment are listed.

*The free energy correction.*

The equation defining the correction potential $\varphi$ (eq 16) reads

$$\nabla(\varepsilon \nabla \varphi) = -4\pi \delta\rho \qquad (B1)$$

The quantity $\delta\rho$, being strongly position – dependent, does not provide a transparent visualization of how significant are the effects appearing beyond the SBCM. More appropriate is the corresponding free energy change $\delta G_{el}$, the integral characteristic. With some manipulations for the standard electrostatic free energy $G_{el} = \frac{1}{8\pi}\int ED d^3r$ (similar to those made at the end of the section 2), the following exact expression can be derived:

$$G_{el} = -\frac{1}{8\pi}\int E_1 \nabla \psi_0 d^3r + \frac{1}{8\pi}\int \nabla\varphi \nabla \psi_0 d^3r \qquad (B2)$$

The first term represents the SBCM energy. The correction, given by the second term, can be estimated in terms of the Green's theorem:

$$\delta G_{el} = -\frac{1}{8\pi}\int \varphi \nabla^2 \psi_0 d^3r = \frac{1}{2}\int \varphi\rho d^3r \qquad (B3)$$

Here $\rho$ is the solute charge distribution.

*Evaluation of $\delta\rho$.*

The quantity $\delta\rho$ (13) is the correction term which is added to the polarization charge density $g(r)$ appearing in accord with eq 5 in the SBCM-2. Its explicit expression is based on the relation $E_2 = \nabla \times A$, where the vector potential $A$ is given in (9). The intermediate result is then obtained as

$$\delta\rho = \frac{1}{4\pi}\nabla\varepsilon\int\left(\nabla\frac{1}{|r-r'|}\right)\times I(r')d^3r'$$

The identities $\nabla(\varepsilon E_2) = \nabla\varepsilon E_2$ and $\nabla \times \left(\frac{I(r')}{|r-r'|}\right) = \nabla \frac{1}{|r-r'|} \times I(r')$ were implemented. The latter one appeared because the operator $\nabla$ acts on the $r$–variable, whereas $I(r')$ depends on $r'$. With $I(r')$ given by (11) we find

$$\delta\rho = \frac{1}{(4\pi)^2} \nabla\varepsilon \int \left(\nabla \frac{1}{|r-r'|}\right) \times \left(\frac{1}{\varepsilon^2(r')} [\nabla'\varepsilon(r') \times \nabla'\psi_0(r')]\right) d^3r' \tag{B4}$$

*The perturbation expansion*

The rearrangement of eq B1 results in the equation

$$\nabla^2 \varphi = -4\pi \frac{\delta\rho}{\varepsilon} - \frac{\nabla\varepsilon \nabla\varphi}{\varepsilon} \tag{B5}$$

Here the first term on the right hand part represents the polarization charge appearing in the bulk solvent beyond the solute cavity. It is therefore scaled by the screening factor $\frac{1}{\varepsilon}$. The second term represents the secondary polarization effect induced by the first term; it is quite similar to the original SBCM charge $g(r)$, introduced in eq. 5. This is seen when the SBCM approximation $\nabla\varphi \approx \frac{1}{\varepsilon}\nabla\overline{\varphi}_0$ is applied in (B5) with $\nabla^2\overline{\varphi}_0 = -4\pi\delta\rho$.

The integral equation following from (B5) reads

$$\varphi(r) = \varphi_0(r) + \varphi_1(r)$$
$$\varphi_0(r) = \int \frac{d^3r'}{|r-r'|} \frac{\delta\rho(r')}{\varepsilon(r')} \tag{B6}$$
$$\varphi_1(r) = \int \frac{d^3r'}{|r-r'|} \frac{\nabla'\varepsilon(r')\nabla'\varphi(r')}{\varepsilon(r')}$$

It gives rise to the standard iterative scheme (the perturbation expansion). As the first step the approximation $\varphi(r) \approx \varphi_0(r)$ is substituted in the expression for $\varphi_1(r)$; it produces the approximate but explicit result $\widetilde{\varphi}_1(r)$ and $\varphi(r) \cong \varphi_0(r) + \widetilde{\varphi}_1(r)$. This approximation, being inserted in the expression for $\varphi_1(r)$, produces the second order correction, etc. The alternative expansion arises when the SBCM approximation is used in the $\varphi_1$ expression, i.e. $\nabla \varphi(r) \cong \frac{1}{\varepsilon} \nabla \overline{\varphi}_0(r)$, where $\overline{\varphi}_0(r)$ appears if the scaling factor $\frac{1}{\varepsilon(r')}$ is eliminated in the expression for $\varphi_0(r)$, eq. B6.

**Appendix C. The perturbational treatment of the Onsager dipole.**

The details of the treatment of the dipolar system examined in section 4 are considered here. The vector notation for the spherical coordinates $\vec{r}(r,\vartheta,\varphi)$ and $\vec{r}\,'(r',\vartheta',\varphi')$ is used, the Cartesian coordinates being $x, y, z$. The vacuum potential created by the dipole is $\psi_0(\vec{r}) = \frac{mz}{r^3}$. By substituting it in eq B4 we obtain

$$\delta\rho(\vec{r}) = -\frac{m^2}{(4\pi)^2} \int f(\vec{r},\vec{r}\,')d^3r'$$

$$f(\vec{r},\vec{r}\,') = \frac{\sin\vartheta'[\cos\vartheta\sin\vartheta' - \cos\vartheta'\sin\vartheta\cos(\varphi-\varphi')]}{r'^2|\vec{r}-\vec{r}\,'|^3} \frac{d}{dr}\varepsilon(r)\frac{d}{dr'}\left(-\frac{1}{\varepsilon(r')}\right) \quad\quad (C1)$$

$$|\vec{r}-\vec{r}\,'| = (r^2 + r'^2 - 2rr'\cos\mu)^{1/2}$$
$$\cos\mu = \cos\vartheta\cos\vartheta' + \sin\vartheta'\sin\vartheta\cos(\varphi-\varphi')$$

The integral in (C1) depends only on the phase difference $(\varphi-\varphi')$, and we conclude therefore that $\delta\rho(\vec{r})$ is $\varphi$ - independent. By this means the response potential $\varphi_0(\vec{R})$ defined in spherical coordinates $\vec{R}(R,\theta,\Phi)$ becomes $\Phi$ - independent, such that

$$\varphi_0(R,\theta) = -2\pi \int \frac{r^2 dr}{\varepsilon(r)} \sin\vartheta d\vartheta \frac{1}{|\vec{R}-\vec{r}|} \int d^3r' \frac{m^2}{(4\pi)^2} f(\vec{r},\vec{r}\,') \quad\quad (C2)$$

This is the final result of the accurate integration. It involves five explicit quadratures (three in (C1) and the two more in (C2)).

A significant simplification is gained if we totally neglect in (C1) the angular dependence of the denominator, which corresponds to the zero order term of the expansion of the kernel $|\vec{r}-\vec{r}'|^{-3}$ in spherical harmonics. This approximate kernel has the form

$$K_0(r,r') = \frac{1}{r^3}[1+\frac{t}{2}\ln\left(\frac{1+t-\delta}{1+t}\right)] \quad (r>r'; t=\frac{r'}{r})$$
$$K_0(r,r') = \frac{1}{r'^3}[1+\frac{t}{2}\ln\left(\frac{1+t-\delta}{1+t}\right)] \quad (r<r'; t=\frac{r}{r'})$$
(C3)

The details of this expression are given in Appendix D; $\delta$ is the fitting parameter supposed to be small. Then the integral in (C2) is explicitly performed by using the expansion of $1/|\vec{R}-\vec{r}|$ in spherical harmonics [11]:

$$\varphi_0(R,\theta) = -\frac{2m}{9} R \cos A(R)$$
$$A(R) = \int_a^\infty dr' [\int_a^R \left(\frac{r}{R}\right)^3 \frac{c(r,r')}{\varepsilon(r)} dr + \int_R^\infty \frac{c(r,r')}{\varepsilon(r)} dr]$$
(C4)
$$c(r,r') = \frac{d\varepsilon(r)}{dr} \frac{d}{dr'}\left(-\frac{1}{\varepsilon(r')}\right) K_0(r,r')$$

Next the evaluation of the second component $\varphi_1(\vec{R})$ of the potential $\varphi(\vec{R})$ (eq 21) is also available within the approximation C3. It involves the quantity

$$b(R') = \frac{1}{\varepsilon(R')} \frac{d\varepsilon(R')}{dR'} [A(R') + R' \frac{dA}{dR'}]$$
$$\frac{dA}{dR'} = -\frac{3}{R'} \int_a^\infty dr' [\int_a^{R'} \left(\frac{r}{R'}\right)^3 \frac{c(r,r')}{\varepsilon(r)} dr]$$
(C5)

Here $A(R')$ is given in eq C4. The final result includes one extra quadrature:

$$\varphi_1(R,\theta) = -\frac{m}{9}\frac{R}{3} \cos\theta B(R)$$
$$B(R) = \int_0^R \left(\frac{R'}{R}\right)^3 b(R') dR' + \int_R^\infty b(R') dR']$$
(C6)

The total free energy excess is obtained as

$$\delta G_{el} = -\frac{2m^2}{9}(A + \frac{1}{3}B)$$

$$A = \int_a^\infty dr' [\int_a^\infty \frac{c(r,r')}{\varepsilon(r)} dr] \qquad (C7)$$

$$B = \int_a^\infty b(R') dR'$$

The quantities $A$ and $B$ represent the values of the function $A(R)$ (eq C4) and $B(R)$ (eq C6) at $R = 0$.

**Appendix D. The motivation and justification of the approximate expression C3.**

We considered here the approximation for the function $K(\vec{r}, \vec{r}') = |\vec{r} - \vec{r}'|^{-3/2}$ which eliminates its angular dependence. It simplifies significantly the subsequent angular integrations in the expression C2. The result C3 is gained by averaging of this function on the surface of the sphere with radius $r = |\vec{r}|$, thus producing the term

$$K_0(\vec{r}, \vec{r}') = \frac{1}{4\pi} \int_0^{2\pi} d\varphi \int_0^\pi \sin\mu \, d\mu \left(\frac{1}{\sqrt{r^2 + r'^2 - 2rr'\cos\mu}}\right)^3 \qquad (D1)$$

where $r' = |\vec{r}'|$ and $\mu$ is the angle between vectors $\vec{r}, \vec{r}'$. We used next the multipole expansion [11,12]:

$$(r^2 + r'^2 - 2rr'\cos\mu)^{-3/2} = -\left(\frac{1}{r_>}\right)^3 \frac{1}{t - \cos\mu} \sum_{l=1}^\infty l t^{l-1} P_l(\cos\mu) \qquad (D2)$$

where

$$t = \begin{cases} \dfrac{r}{r'} & (r < r') \\ \dfrac{r'}{r} & (r > r') \end{cases} \qquad (D3)$$

and $r_> = \max(r, r')$. The Legendre polynomials of argument $x$ are denoted as $P_l(x)$. The result (D2) can be obtained from the standard expression

$$\frac{1}{|\vec{r}-\vec{r}\,'|}=\frac{1}{r_>}g(t);\quad t<1$$

$$g(t)=\frac{1}{\sqrt{1+t^2-2t\cos\mu}}=\sum_{l=0}^{\infty}t^l P_l(\cos\mu)$$
(D4)

By evaluating the derivative

$$\frac{\partial g(t)}{\partial t}=[g(t)]^3(\cos\mu-t)$$
(D5)

the expansion (D2) is readily reconstructed. Its first term $(l=1)$ is inserted in (D1) to yield the approximation which appears after performing the integral:

$$K_0\cong\frac{1}{r_>^3}\left(1+\frac{t}{2}\ln\frac{1-t+\delta}{1+t}\right);\quad t<1$$
(D6)

Strictly speaking, it is supposed that $\delta\to+0$ in (D6). Thereby, this expression diverges when $r=r'$. The weak singularity is eliminated during the further integrations over $r'$ and $r$ according to eqs 23 and C4-C6. Within the simplified treatment, accepted in our approximate calculations, the quantity $\delta>0$ was considered as an adjustable parameter providing the converged values of $K_0(r,r')$ for $r=r'$.

We consider finally the stepwise limit of the free energy misfit $\delta G_{el}$. It corresponds to the extreme $\alpha\to\infty$ performed in the dielectric function (20). For the exact solution of the Poisson equation 16 this limit represents the PCM approach [10] which reduces to the standard Onsager problem [13] for the special case of our dipolar system. Then the SBCM algorithm can be performed analytically [1] providing the expression:

$$\delta G_{el}=G_{el}(PCM)-G_{el}(SBCM)=-\frac{m^2}{6a^3}\left(1-\frac{1}{\varepsilon_0}\right)\left(\frac{\varepsilon_0-1}{\varepsilon_0+1/2}\right)$$
(D7)

Here $m$ is the dipole moment value, $a$ is the radius of the spherical cavity and $\varepsilon_0$ is the bulk (static) dielectric permittivity (see section 4). This result appears because $\varepsilon(r)$ is the Heaviside step function and $d\varepsilon(r)/dr$ becomes the delta-function in the step limit. The integrals involved in the theory are performed using this property at point $r=a$, whereas the smooth functions entering their integrands are withdrawn out if the integration symbol at this point. The similar manipulation can be applied to the expression C7 with the following outcome: in the step limit we find $B = A \ln \varepsilon_0$ whereas the evaluation of $A$ yields the ultimate result

$$\delta G_{el} = -\frac{m^2}{9}\left(1 - \frac{1}{\varepsilon_0}\right) \ln \varepsilon_0 \left(1 + \frac{\ln \varepsilon_0}{3}\right) K_0(a,a) =$$
$$= -\frac{m^2}{9a^3}\left(1 - \frac{1}{\varepsilon_0}\right) \ln \varepsilon_0 \left(1 + \frac{\ln \varepsilon_0}{3}\right)\left(1 + \frac{1}{2}\ln(\delta/2)\right) \quad (D8)$$

The representation D6 for $K_0$ is invoked here. The spurious logarithmic dependence on $\varepsilon_0$ is the consequence of approximations involved in (D6) (the true singularity at $r = r'$ is cancelled because $\delta$ is treated as a finite positive number). This defect can formally be eliminated by assuming that $\delta$ is a function of $\varepsilon_0$; so the expressions D7 and D8 become identical when

$$\ln(\delta/2) = \frac{3(\varepsilon_0 - 1)}{\ln \varepsilon_0 (1 + \ln \varepsilon_0 /3)(\varepsilon_0 + 1/2)} - 2 \quad (D9)$$

In this way the present approximate model provides the result for $\delta G_{el}$ which turns into the exact non-perturbational value of $\delta G_{el}$ in the step limit $\alpha \to \infty$, when $\delta$ is defined according to eq D9. It can be extrapolated to finite $\alpha$ values in terms of eq C7. The quality of this approach is visualized by Fig.1, where the exact (eq 21 with $\varphi = \varphi_0$ being computed according to eqs C1, C2) and approximate (eq 24) calculations for the zero-order term $\delta G_{el}^{(0)}$ are made for $\varepsilon_0 = 20$ with $\delta(20) = 0.43$, as extracted from the equation D9.

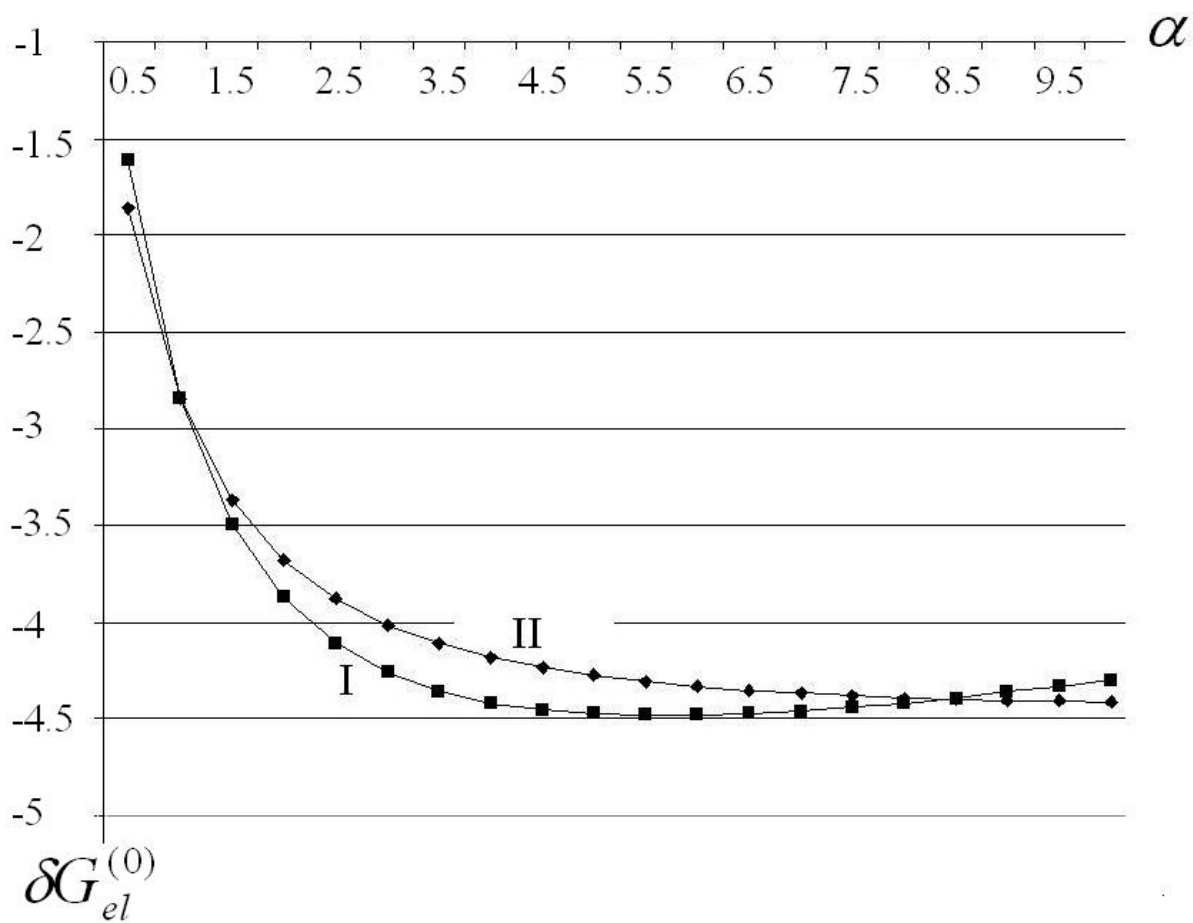

Figure 1. The zero-order energy misfit $\delta G_{el}^{(0)}$ (kcal/mol) as a function of the steepness parameter $\alpha$ (Å$^{-1}$): I- the rigorous computation (eqs C1, C2); II- the approximate computation (eqs 23, C3). The parameters are: $m = 4.8 D$, $\varepsilon_0 = 20$, $a = 2$ Å, $\delta = 0.43$ (eq 29).

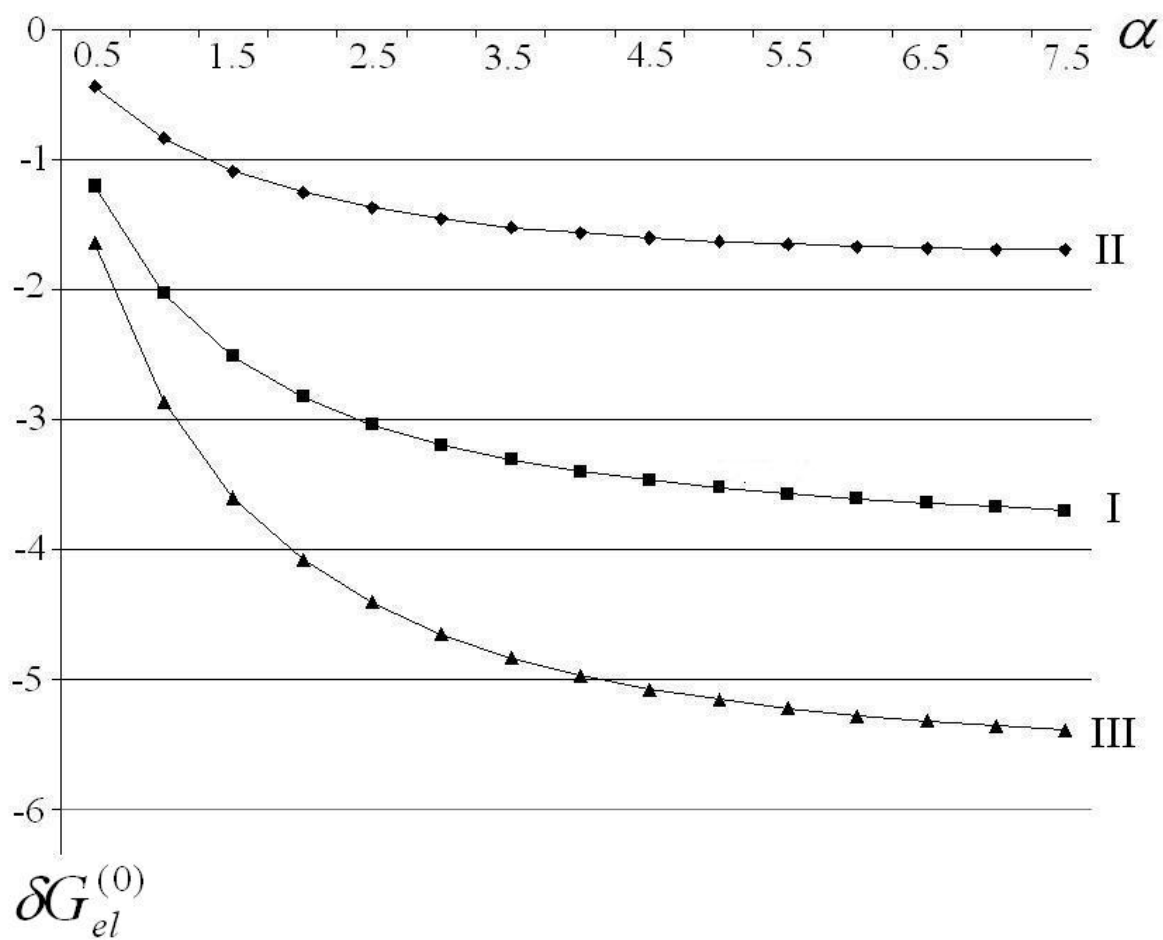

Figure 2. The comparison of the zero-order $\delta G_{el}^{(0)}$ (I) and first-order $\delta G_{el}^{(1)}$ (II) components of the free energy misfit $\delta G_{el}$ (III) as a function of $\alpha$ (Å$^{-1}$). The approximate computation (eqs 23 and C3) was performed with parameters $m = 4.8 D$, $\varepsilon_0 = 20$, $a = 2$ Å, $\delta = 0.43$ (eq 29). Energies are given in kcal/mol.

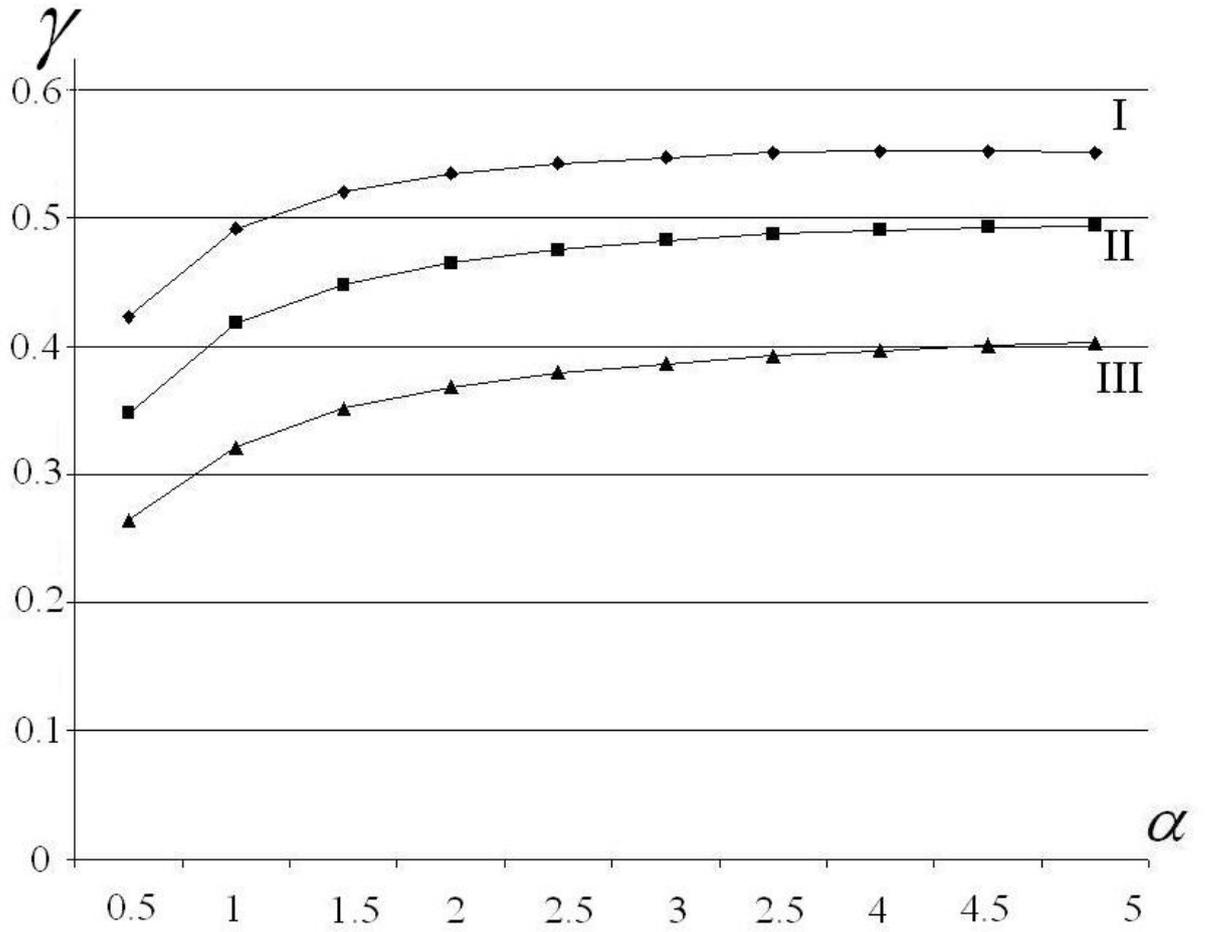

Figure 3. The dimensionless misfit parameter $\gamma$ (eq 26) as a function of $\alpha$ (Å$^{-1}$): $\varepsilon_0 = 20$, $\delta = 0.43$ (I); : $\varepsilon_0 = 10$, $\delta = 0.51$ (II); : $\varepsilon_0 = 5$, $\delta = 0.65$ (III). The cavity radius is $a = 2$ Å.

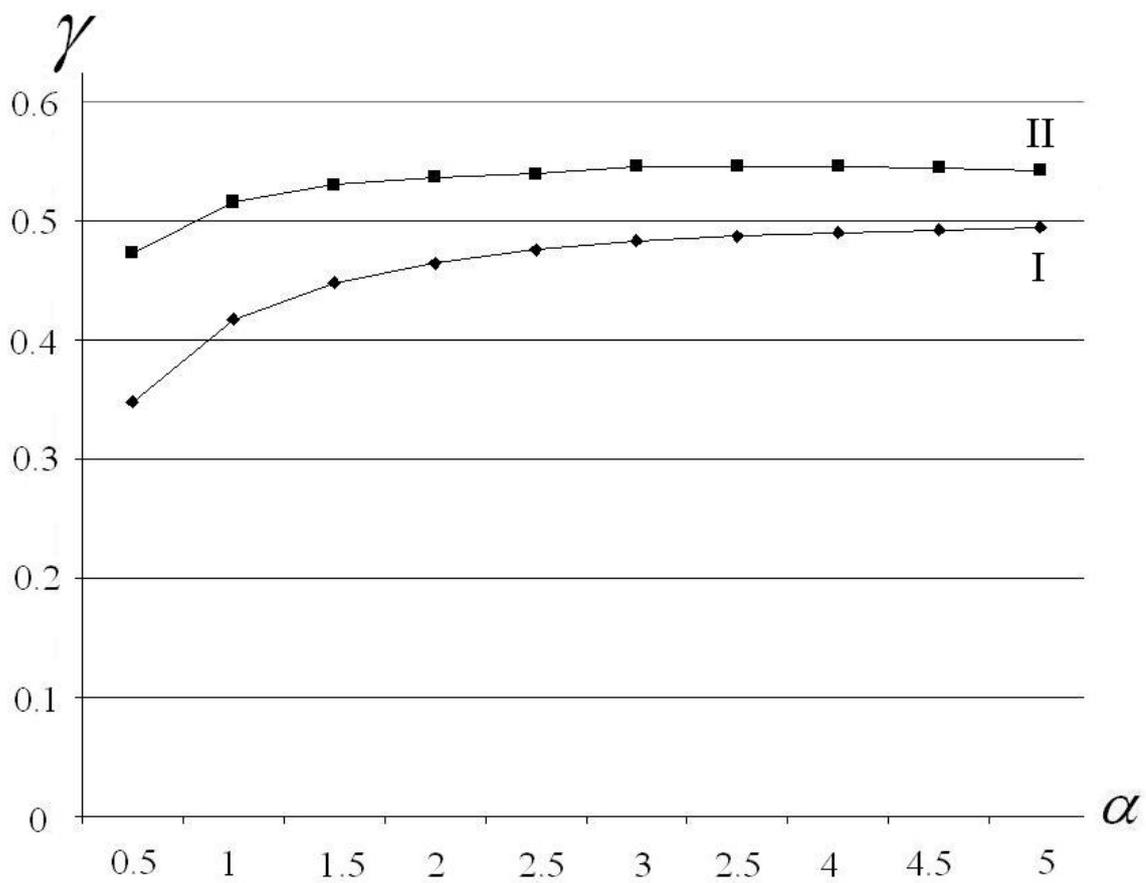

Figure 4. The dependence of the misfit parameter $\gamma$ on the cavity radius $a$ (Å): $a = 2$ (curve I); $a = 3$ (curve II); $\varepsilon_0 = 10$ for both cases.

**Figure captions**

Figure 1. The zero-order energy misfit $\delta G_{el}^{(0)}$ (kcal/mol) as a function of the steepness parameter $\alpha$ (Å$^{-1}$): I- the rigorous computation (eqs C1, C2); II- the approximate computation (eqs 23, C3). The parameters are: $m = 4.8 D$, $\varepsilon_0 = 20$, $a = 2$ Å, $\delta = 0.43$ (eq 29).

Figure 2. The comparison of the zero-order $\delta G_{el}^{(0)}$ (I) and first-order $\delta G_{el}^{(1)}$ (II) components of the free energy misfit $\delta G_{el}$ (III) as a functions of $\alpha$ (Å$^{-1}$). The approximate computation (eqs 23 and C3) was performed with parameters $m = 4.8 D$, $\varepsilon_0 = 20$, $a = 2$ Å, $\delta = 0.43$ (eq 29). Energies are given in kcal/mol.

Figure 3. The dimensionless misfit parameter $\gamma$ (eq 26) as a function of $\alpha$ (Å$^{-1}$): $\varepsilon_0 = 20$, $\delta = 0.43$ (I); : $\varepsilon_0 = 10$, $\delta = 0.51$ (II); : $\varepsilon_0 = 5$, $\delta = 0.65$ (III). The cavity radius is $a = 2$ Å.

Figure 4. The dependence of the misfit parameter $\gamma$ on the cavity radius $a$ (Å): $a = 2$ (curve I); $a = 3$ (curve II); $\varepsilon_0 = 10$ for both cases.